# Evidence of a topological Hall effect in Eu$_{1-x}$Sm$_x$TiO$_3$

**Kaveh Ahadi[a)], Luca Galletti, and Susanne Stemmer**

Materials Department, University of California, Santa Barbara, California 93106-5050, USA

[a)] Corresponding author.  Email: kaveh.ahadii@gmail.com

**Abstract**

We report on the observation of a possible topological Hall effect in thin films of the itinerant ferromagnet $Eu_{1-x}Sm_xTiO_3$. $EuTiO_3$ and $Eu_{0.955}Sm_{0.045}TiO_3$ films were grown by molecular beam epitaxy. The $EuTiO_3$ film is insulating. The Hall resistivity of the $Eu_{0.955}Sm_{0.045}TiO_3$ films exhibits the anomalous Hall effect below the Curie temperature of ~ 5 K as well as additional features that appear at 2 K. It is shown that these features are magnetic in origin and consistent with the topological Hall effect seen in materials systems with topologically nontrivial spin textures, such as skyrmions. The results open up interesting possibilities for epitaxial hybrid heterostructures that combine topological magnetic states, tunable carrier densities, and other phenomena.

The topological Hall effect (THE) is a hallmark of topologically nontrivial (chiral) spin textures, such as skyrmions, and can be observed as a distinct, additional contribution in Hall measurements that is superposed on the ordinary and anomalous Hall effects [1,2]. It arises from the "Berry phase" that the conduction electrons acquire when moving in the topologically protected spin texture [3-5]. Skyrmions are a result of the Dzyaloshinskii-Moriya interaction in chiral magnets with a non-centrosymmetric crystal structure, such as MnSi [6], and they have generated significant interest for application in nonvolatile memories [2]. The topological Hall effect has also been observed in centrosymmetric crystals, such as $SrFe_{1-x}Co_xO_3$ and pyrochlores [7,8], where helical structures form as a result of magnetic frustration, and in thin films, where inversion symmetry may be lifted by interfaces and surfaces [9,10]. Oxide films and interfaces that support topologically nontrivial spin textures and whose carrier density can be manipulated are interesting, because the potential for control by electric field effect and because proximity effects can be utilized to realize other exotic states within all-epitaxial heterostructures.

Doped $EuTiO_3$ films are attractive for such approaches. In its stoichiometric form, $EuTiO_3$ is a quantum paraelectric with the cubic perovskite structure at room temperature, similar to $SrTiO_3$ [11,12]. The Eu magnetic moments [$4f^7$ ($S = 7/2$)] order in a G-type antiferromagnetic pattern below the Neel temperature of 5.5 K [13-15]. Strained $EuTiO_3$ films can become simultaneously ferromagnetic and ferroelectric at low temperatures [16]. Chemical doping, i.e. substitution of $Eu^{+2}$ with a trivalent rare earth ion, introduces itinerant electrons into the Ti $3d$ $t_{2g}$ derived conduction band states. The material then becomes a ferromagnetic metal due to Ruderman-Kittel-Kasuya-Yoshida (RKKY) type interactions between the localized 4f and itinerant 3d $t_{2g}$ electrons [17,18]. La-doped $EuTiO_3$ films exhibit a strong anomalous Hall effect (AHE) whose sign can be manipulated by the carrier density [18]. Here, we investigate $Eu_{1-x}Sm_xTiO_3$ thin films grown by hybrid molecular beam epitaxy (MBE). We show that in addition to a positive AHE, a topological Hall effect (THE) component appears in the Hall measurements at temperatures below 5 K.

Epitaxial, 50-nm-thick $Eu_{1-x}Sm_xTiO_3$ layers were grown by MBE on (001) $(La_{0.3}Sr_{0.7})(Al_{0.65}Ta_{0.35})O_3$ (LSAT) single crystals. Elemental Eu and a metalorganic, titanium tetra isopropoxide (TTIP), were used to supply Eu, Ti, and oxygen, respectively. The MBE approach is similar to that previously used for $SrTiO_3$ [19,20], except that no extra oxygen was supplied. X-ray diffraction (XRD) showed that the films were single-phase and thickness fringes indicated smooth films of high structural quality (see Supplementary Information). Electron beam evaporation through a shadow mask was used to deposit Au/Ti (400/40 nm) contacts for Hall and longitudinal (magneto-)resistance measurements using square Van der Pauw structures that were contacted via Au wire bonds. Temperature ($T$) dependent magnetotransport measurements were carried out using a Quantum Design Physical Property Measurement System (PPMS) with excitations currents of 200 µA. A dilution refrigerator was used for some of the data shown in the Supplementary Information.

Figure 1(a) shows the sheet resistance $R_s$ as a function of temperature for $Eu_{1-x}Sm_xTiO_3$ thin films with $x = 0$ and $x = 0.045$, respectively. The nominally undoped $EuTiO_3$ film is highly resistive and quickly exceeds the measurement limit below room temperature while the doped sample shows metallic ($dR_s/dT > 0$) behavior. Figure 1(b) shows the temperature dependence of the carrier concentration for the $Eu_{0.955}Sm_{0.045}TiO_3$ film as determined from the ordinary Hall effect. The room temperature carrier concentration is $9\times10^{20}$ cm$^{-3}$, close to the expected carrier density ($8\times10^{20}$ cm$^{-3}$) from the nominal doping concentration. It shows some temperature dependence, which may indicate some carrier trapping but can also be a result of the particular behavior of Hall and resistance scattering rates, typical for these oxides, as discussed elsewhere [21,22].

Figure 2(a) shows the temperature dependence of $R_s$ under various magnetic fields ($B$) applied normal to the film ($B = 0$, 1, 3, 6, and 9 T) at low temperatures (2 K – 50 K) for the $Eu_{0.955}Sm_{0.045}TiO_3$ film. Without $B$, an upturn in $R_s$ emerges at 20 K followed by a sharp drop at ~ 6 K. The magnetic field suppresses the upturn systematically until it almost vanishes at 9 T. Similar behavior was reported previously for La [17] and Nb [23]-doped $EuTiO_3$ and is due to the alignment of the localized spins under the magnetic field, which reduces the resistance. Figure 2(b) shows $R$ vs. $B$ (magnetoresistance) for the $Eu_{0.955}Sm_{0.045}TiO_3$ film at 2 K, 5 K, 10 K and 15 K, respectively. Here, $B$ was swept from -9 T to +9 T and back. All samples show negative magnetoresistance, which peaks at 5 K, consistent with Fig. 2(a). Hysteresis in the magnetoresistance can be noticed at 5 K and below [see also Fig. 2(c)]. The hysteresis confirms the ferromagnetism reported in the literature [17,18].

Figures 3(a,b) show the Hall results for the $Eu_{0.955}Sm_{0.045}TiO_3$ film at $T = 2$ K (a) and 5 K (b), respectively. The Hall data was antisymmetrized to eliminate the magnetoresistance contribution, i.e., $R_{xy} = [R_{xy}^{raw}(+B) - R_{xy}^{raw}(-B)]/2$, where the superscript indicates the raw data. A linear fit to $R_{xy}$ at high $B$ (6 T – 9 T) yields the ordinary Hall component ($R_0B$) and is also shown. Comparing the two curves shows that at low fields $R_{xy}$ deviates from linearity. In the presence of an AHE and/or a THE, $R_{xy}$ is given by: $R_{xy} = R_0H + R_{AHE} + R_{THE}$, where $R_{AHE}$ and $R_{THE}$ are the anomalous and topological effects, respectively [1,2]. Figures 3(c) and (d) show the data after subtraction of $R_0B$ from $R_{xy}$ at $T = 2$ K (c) and additional temperatures (d), respectively. At 5 K and above, the monotonic increase in the resistivity with $B$ is characteristic for the conventional AHE, described as $R_{AHE} = \alpha M R_{xx0} + \beta M R_{xx0}^2 + \gamma M R_{xx}^2$, where $\alpha, \beta,$ and $\gamma$ correspond to skew scattering, side jump, and intrinsic components, respectively, and $R_{xx0}$ is the residual resistance [24]. In these films it is expected that the high resistivity (~ $10^{-4}$ $\Omega$cm) causes the AHE to be dominated by the intrinsic, Berry phase contribution [25].

At 2 K [Fig. 3(c)] peaks appear in addition to the AHE. The peak at ~0.8 T (stronger in upward than downward sweep) is near the field at which the closure of the hysteresis is seen in the magnetoresistance data shown in Fig. 2(b). A second, more pronounced peak at ~ 2 T has no corresponding feature in the magnetoresistance. These general features are very similar to the THE signal found in a wide range of other materials with magnetic skyrmions [1,2,9]. Their origin is magnetic as the peaks only appear below the Curie temperature. Therefore, nonmagnetic phenomena like multicarrier Hall response cannot be the origin. Furthermore, they can be suppressed with an in-plane field (see Supplementary Information), which indicates that a two-dimensional spin texture gives rise to the THE [9]. It is interesting that the THE only appears at 2 K, whereas the AHE and magnetoresistance hysteresis are already pronounced at 5 K, which may indicate a possible phase transition to a topological non-trivial spin structure. Preliminary measurements at 40 mK indicate that the features assigned here to the THE persist to very low temperatures (see Supplementary Information). This indicates that the THE can be associated with a non-collinear structure. The persistence to low temperatures can be contrasted with that of helical MnSi, where thermal fluctuations are required to stabilize the skyrmion state [6]. It is important to note, however, that chiral magnetic textures with zero topological charge can exhibit Hall effects that are quite similar to the THE of magnetic skyrmions [26,27]. Therefore, future experiments should focus on an independent measurement that could confirm a topologically non-trivial spin structure.

In conclusion, undoped and Sm-doped EuTiO$_3$ films were grown by MBE. The Sm-doped sample is an itinerant ferromagnet below 5 K that exhibits the AHE. At 2 K, signatures of a THE are observed that are indicative of a non-trivial spin structure. Further investigations of the spin texture and determining the space group of compressively strained films, which may be non-centrosymmetric [16], should allow for insights into the origin of the spin structures that give rise to the THE. The results open a wealth of possibilities for future investigation. Because the carrier density can be tuned, this may allow for studies of a quantized THE [28]. Interfaces with other rare earth titanates may allow for interfacial electron systems due to the polar discontinuity, in analogy with GdTiO$_3$/SrTiO$_3$ and SmTiO$_3$/SrTiO$_3$ [29], introducing charge carriers without chemical doping and reducing disorder. Finally, EuTiO$_3$ is perfectly lattice matched with SrTiO$_3$, which can be superconducting, and their combination in epitaxial structures may be of interest for novel superconducting states [30].

See supplementary material [link to be inserted by publisher] for XRD data, measurements of the THE as a function of the magnetic field orientation and at 40 mK.

We acknowledge support from the U.S. Army Research Office (W911NF-14-1-0379) and FAME, one of six centers of STARnet, a Semiconductor Research Corporation program sponsored by MARCO and DARPA. The dilution fridge used in the measurements was funded through the Major Research Instrumentation program of the U.S. National Science Foundation (award no. DMR 1531389). The work made also use of the Central facilities supported by the MRSEC Program of the U.S. National Science Foundation under Award No. DMR 1121053.

# Figure Captions

**Figure 1:** (a) Temperature dependence of $R_s$ of Eu$_{1-x}$Sm$_x$TiO$_3$ films ($x = 0$ and $x = 0.045$). (b) Temperature dependence of Hall carrier concentration ($n$) for the Eu$_{0.955}$Sm$_{0.045}$TiO$_3$ film.

**Figure 2:** (a) Temperature dependence of $R_s$ near the Curie temperature for the Eu$_{0.955}$Sm$_{0.045}$TiO$_3$ film under various applied $B$ fields. (b) Magnetoresistance for the Eu$_{0.955}$Sm$_{0.045}$TiO$_3$ film at different temperatures. (c) Same data at 2 K and 15 K as in (b) but on a different scale to show the hysteresis. The resistance data for the 2 K measurement is the left axis and for the 15 K measurement is the right axis. The arrows indicate the sweep direction of $B$.

**Figure 3:** (a-b) $R_{xy}$ at 2 and 5 K, respectively, and a linear fit that shows the ordinary Hall component. (c) Same data as in (a-b) but after subtraction of the ordinary, linear Hall component ($R_0B$). (d) THE and AHE at different temperatures.

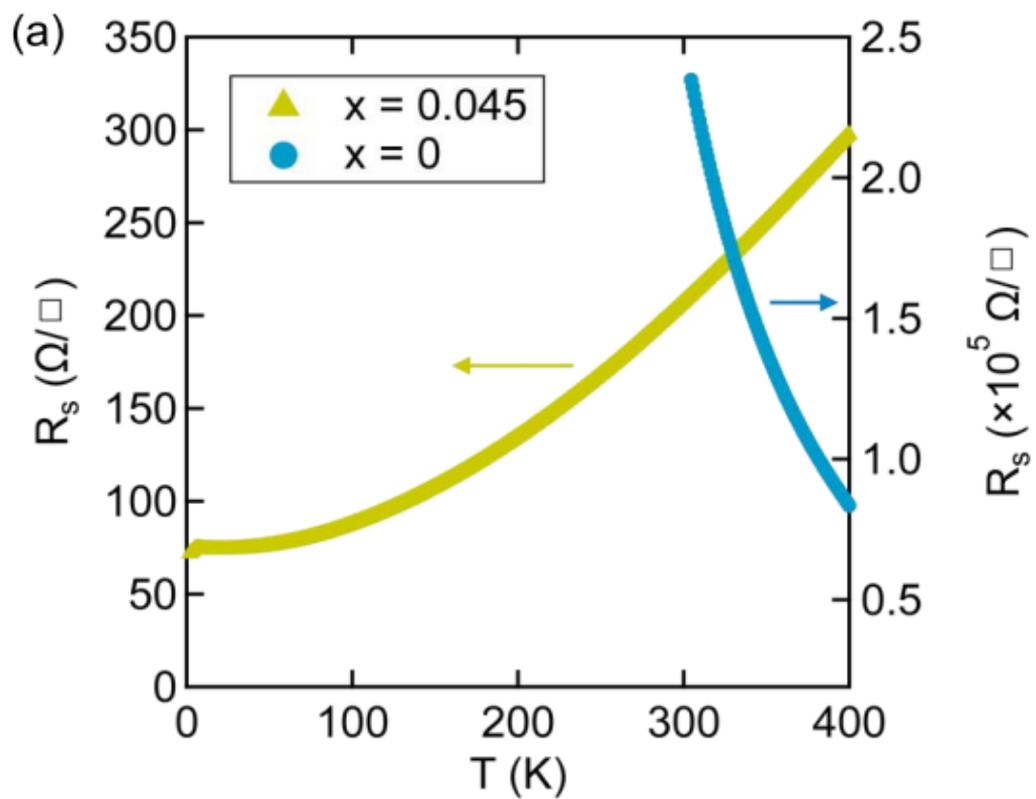

(a)

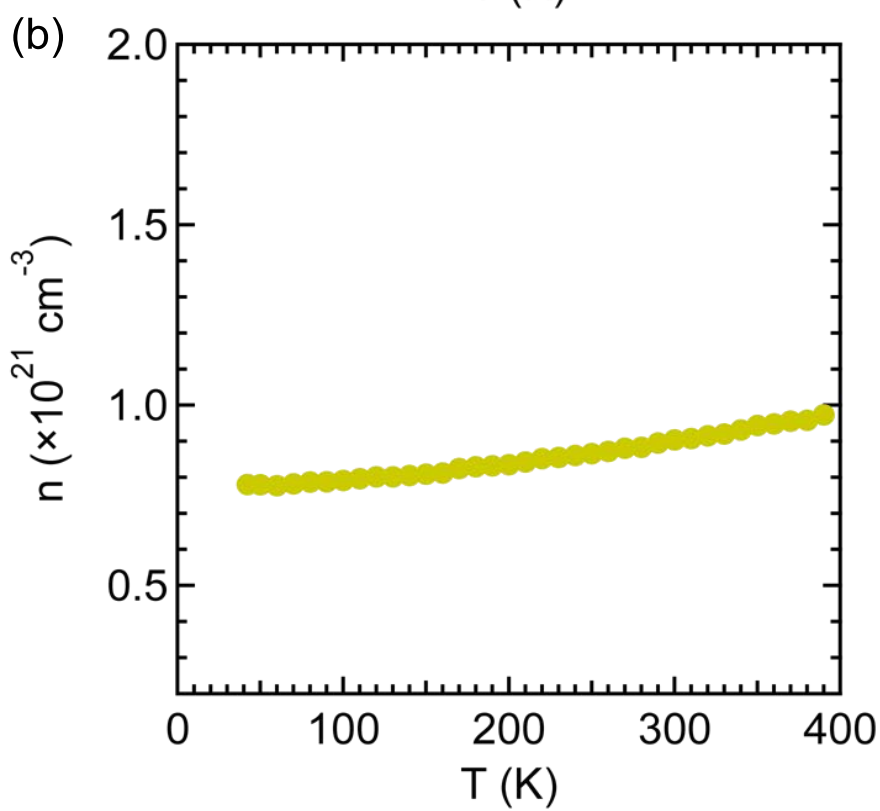

(b)

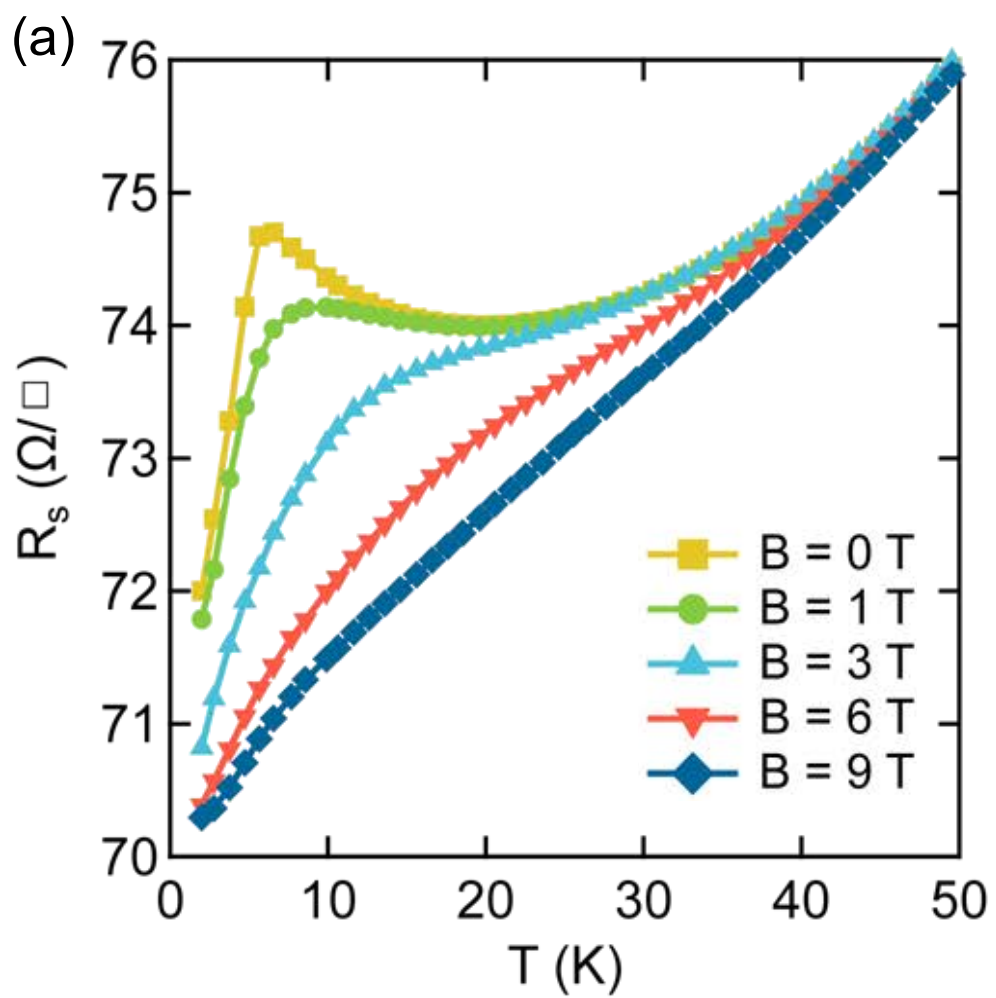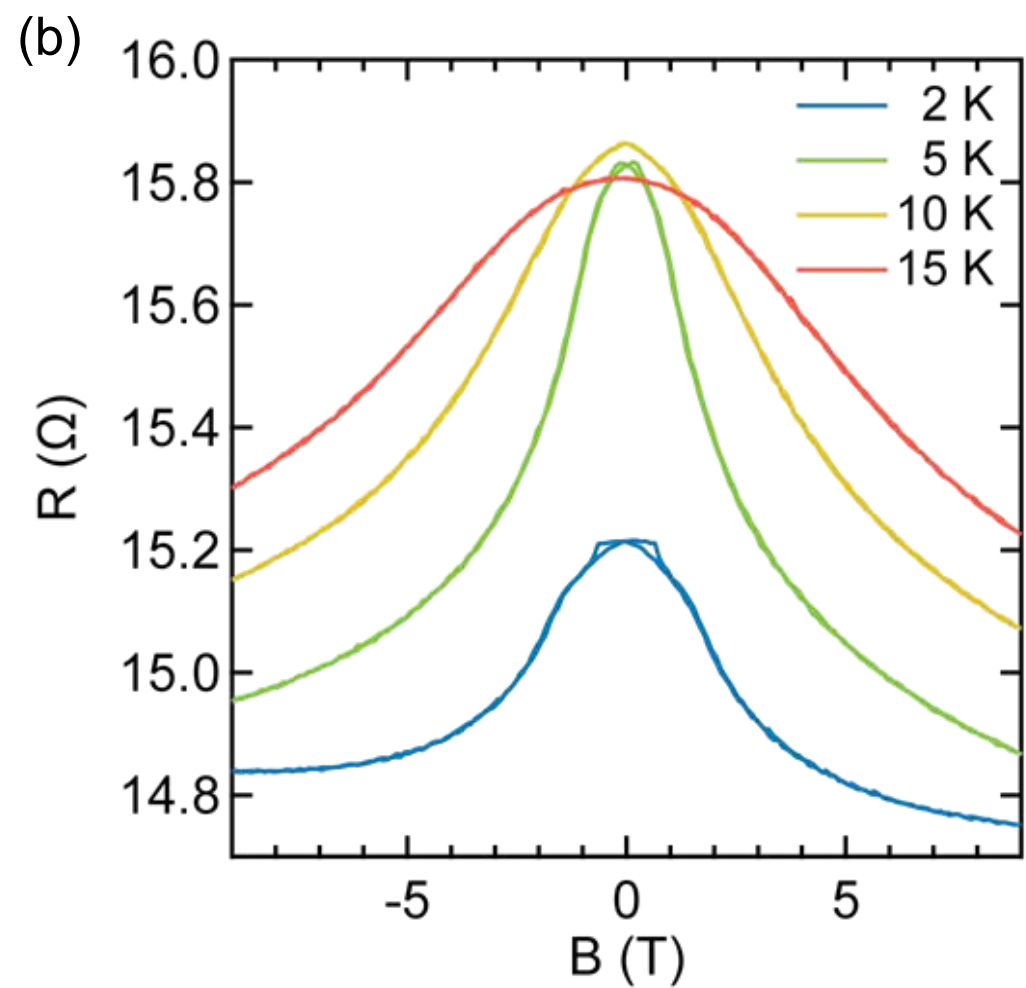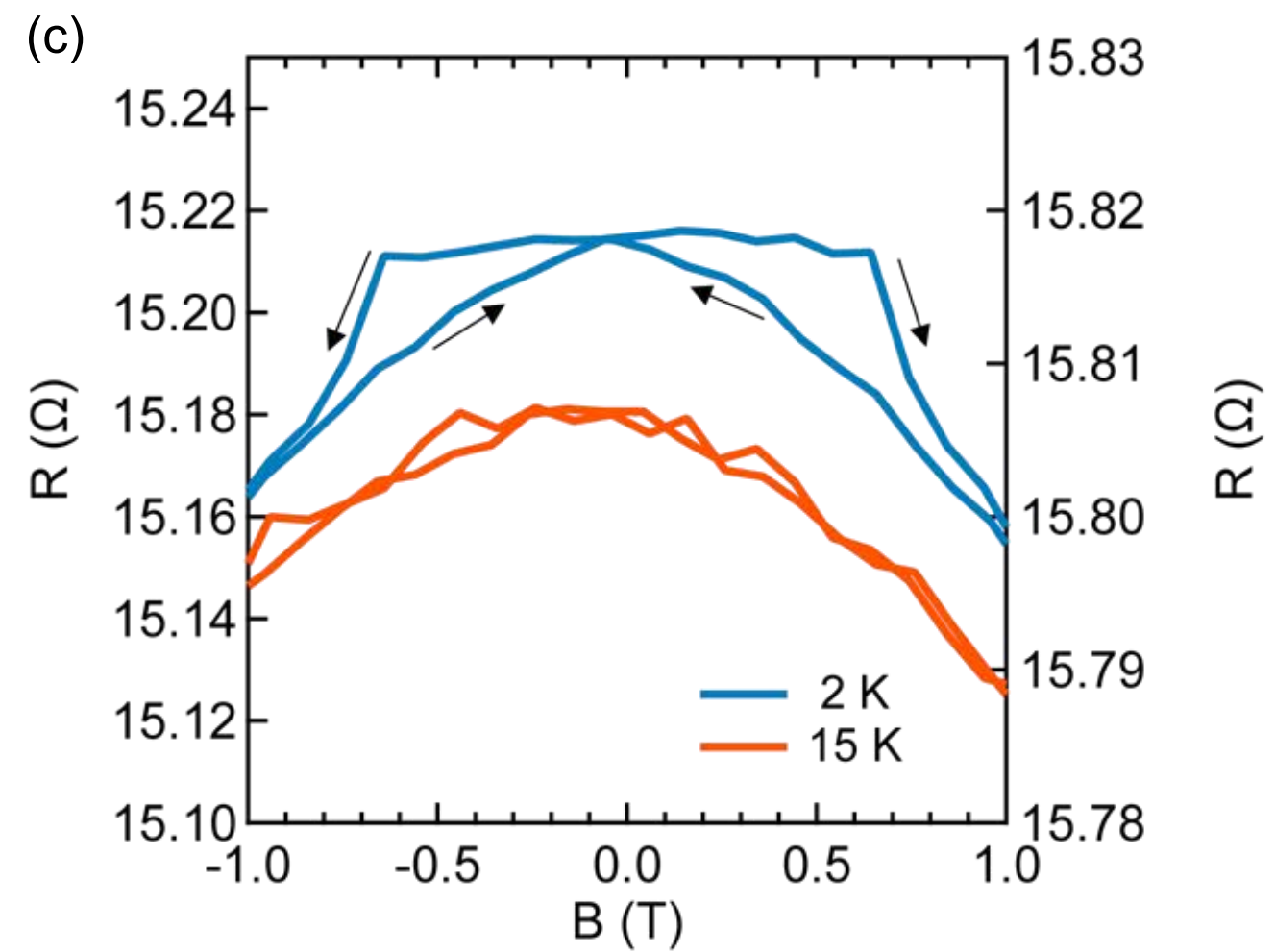

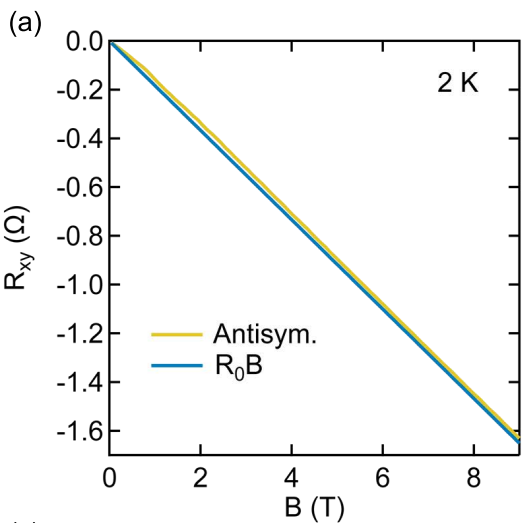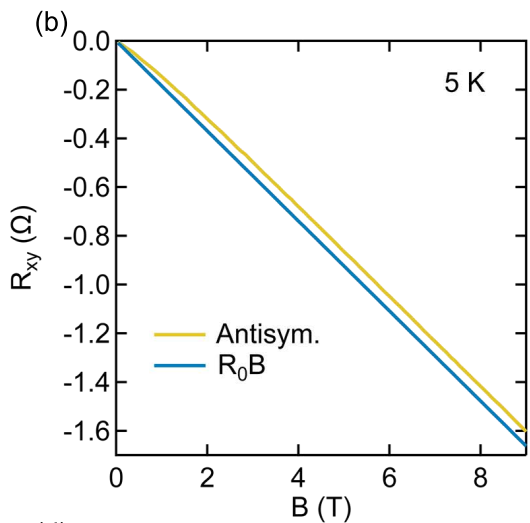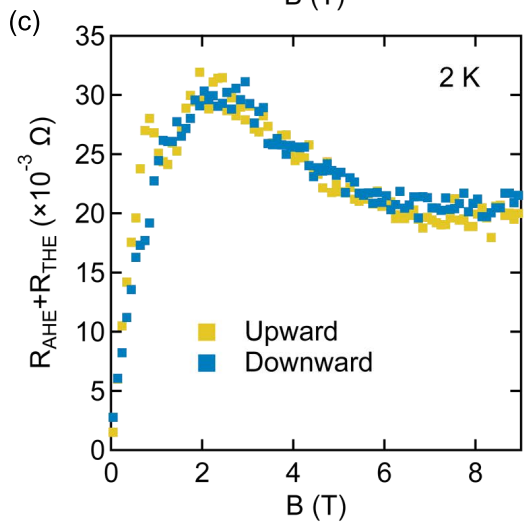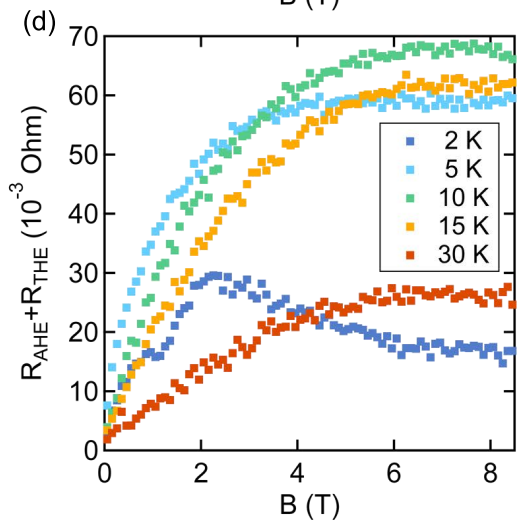